# Ferrofluid Capillary Rise in Porous Medium under the Action of Non-Uniform Magnetic Field

**Arthur Zakinyan** [1]
Department of General and Theoretical Physics, Institute of Mathematics and Natural Sciences, North Caucasus Federal University
1 Pushkin Street, Stavropol 355009, Russian Federation
e-mail: zakinyan.a.r@mail.ru

**Levon Mkrtchyan**
Department of General and Theoretical Physics, Institute of Mathematics and Natural Sciences, North Caucasus Federal University
1 Pushkin Street, Stavropol 355009, Russian Federation

**Victoria Grunenko**
Department of General and Theoretical Physics, Institute of Mathematics and Natural Sciences, North Caucasus Federal University
1 Pushkin Street, Stavropol 355009, Russian Federation

**Yuri Dikansky**
Department of General and Theoretical Physics, Institute of Mathematics and Natural Sciences, North Caucasus Federal University
1 Pushkin Street, Stavropol 355009, Russian Federation

## ABSTRACT

*The capillary flow of a ferrofluid in a single cylindrical capillary tube and through a sandy porous medium under the action of a non-uniform magnetic field is studied experimentally. The dynamics of the capillary rise and the static case have been considered. It has been shown that the non-uniform magnetic field with upward directed gradient accelerates the capillary rise; contrary, the non-uniform magnetic field with downward directed gradient decelerates the capillary rise. Time dependences of the ferrofluid height and maximum reachable height of ferrofluid have been analyzed. The method of the study of ferrofluid capillary rise based on the use of magnetic measurements has been proposed. It has been demonstrated that porous*

---

[1] Corresponding author

*material parameters can be extracted from the results of measurements of the inductances of the solenoid with porous medium inside and the small sensing coil within a single experiment.*

## INTRODUCTION

Capillary driven flow is an important field of research because it has many applications in science, industry and daily life [1–3]. Fluid flows in porous media are strongly influenced by capillarity effects. Examples where fluid flow in porous media plays an important role are drainage and irrigation of agricultural lands, groundwater movement, nuclear waste storage, oil and gas winning, drying of food materials, heat pipes, spacecraft propellant management devices, marker pens, etc. [4]. Investigations of a ferrofluid flow through the porous media have found some new features connected with the capability of influence of external magnetic fields [5]. A ferrofluid (or magnetic fluid) is a colloidal suspension of ultra-fine ferro- or ferrimagnetic nanoparticles suspended in a carrier fluid. Because of the small size of the ferromagnetic particles, ferrofluids can flow through porous media such as natural sediments or fractured rock due to gravitational, pressure gradient, capillary, and magnetic forces [6]. The ferrofluid flow can be manipulated through control of the external magnetic field and without any direct physical contact. Controllable fluid motion in porous media (which are nonmagnetic), without direct access to the fluid, may have the significant applications, such as controlled emplacement of liquids or treatment chemicals, subsurface environmental engineering applications, etc. This has motivated recent interest in the study of ferrofluid capillary rise in porous media.

For the better understanding of the mechanisms of this flow control by the magnetic field, it is important to study ferrofluid behavior in a separate capillary tube without any porous material. Previously several aspects of ferrofluid motions in a separate capillary tube have been considered [7–9]. The stationary height (as a characteristic of capillary force) of a ferrofluid in a separate capillary under the action of uniform magnetic field has been analyzed in [7, 8]. The effect of the uniform magnetic field on the dynamics of magnetic fluid ascension into a vertical cylindrical capillary has been analyzed in [9]. It was found that due to the fluid meniscus deformation, the surface pressure drop decreases in the magnetic field (directed transverse to the capillary) leading to the decrease in the height and the velocity of the capillary ascension. The horizontal and vertical displacements of some volume of ferrofluid in sandy porous media under the action of lateral magnetic force produced by a permanent magnet have been studied in [10, 11]. It was shown that gravitational and magnetic forces result in a predictable configuration of the fluid in the porous medium. Different aspects of ferrofluid motions in porous media have been considered in theoretical studies [12–19]. The capillary ascension of ferrofluid in a porous medium has not been studied previously. The capillary ascension of ferrofluid affected by the non-uniform magnetic field producing the net force on the ferrofluid has also not been studied in the existing works. The work presented here extends the previous works by considering the dynamics of ferrofluid capillary rise in porous media under the action of non-uniform magnetic fields. The linear capillary transport and capillary rise experiments are commonly performed in cylindrical vertical columns filled with porous material, using visual methods to register the advance of the wetting front [20, 21]. We will follow such methodology in our study.

## MATERIALS AND METHODS

The experimental setup used in our experiments is shown in Fig. 1. It consists of a glass rectangular reservoir filled with ferrofluid. A stable ferrofluid level in the reservoir was kept by the overhead tank (not shown in Fig. 1). First we examined the ferrofluid rise in a single capillary. To do this, the cylindrical glass tube with the inner diameter of 200 μm was inserted into a reservoir with ferrofluid. The glass tube was supplied by OJSC "THERMOPRIBOR" (Russia). The tube was installed on a support mount (not shown in Fig. 1) and vertically oriented. The tube was also labeled with a millimeter scale. Once the bottom of the capillary tube contacted with ferrofluid surface, the capillary rise of ferrofluid was observed and recorded by a high-speed digital camera (Casio EXILIM EX-F1). An analogy exists between linear capillary transport in a cylindrical tube and linear capillary transport in a porous structure. The advantage of the capillary tube is that (in contrast to porous media) all of its corresponding pore structure parameters are known with the tube radius and the porosity equal to unity. Thus, as the physical processes occurring in both systems are the same, the capillary tubes can be considered as a good model medium for theoretical and experimental investigations of the flow in other porous structures. To study the ferrofluid rise in porous medium we used a cylindrical glass tube with the inner diameter of 5 mm filled with sand and inserted into a reservoir with ferrofluid. The length of the tube was larger than the maximum reachable fluid height. We used fine deposited river sand with a low content of organic matter provided by local supplier. Sand grain diameters were ranged from 0.03 to 0.3 mm. The microscopic image of the grains of the sand sample used in the experiments is shown in Fig. 2. Sand samples were cleaned using the distilled water, dried in an oven at 105 °C and then cooled at room temperature. A moving interface formed between the wet and non-wet regions of the sandy porous medium was observed and recorded by a digital camera.

An electromagnet driven by a dc current source has provided a non-uniform magnetic field. To produce the non-uniform magnetic field with downward directed gradient the electromagnet poles were placed at the bottom of the glass tube (Fig. 1). To produce the upward directed magnetic field gradient the electromagnet poles were placed at the top of the glass tube. The tube has been mounted along the centerline between the electromagnet poles. The external magnetic field strength $H$ produced by the electromagnet has been measured along the glass tube. It has been revealed that the magnetic field variation with height should only be considered. For the tube centrally located between the two poles, the magnetic field change within the horizontal cross section of a glass tube can be neglected. The measurements have shown that the magnetic field changes roughly linearly with height. For the simplicity of calculations and experimental results analysis we will assume that the magnetic field gradient in vertical direction is constant and its absolute value is equal to $10^6$ A/m$^2$. The corresponding results of measurements of the magnetic field strength as a function of the distance from the electromagnet poles are shown in Fig. 3.

The fact that a ferrofluid has magnetic properties makes it possible to magnetically sense flow in porous media. To this end we used a solenoid with the sand inside instead of just a glass tube. The solenoid inner diameter was 5 mm and its length was 6.5 cm.

The solenoid length is large compared with the diameter and at the same time less than maximum reachable (static) fluid height. So the ferrofluid can reach the top of the solenoid during the ascension process. The solenoid was connected to the measuring bridge to measure the inductance (Fig. 1). In our experiments we used the measuring bridge "GW Instek LCR meter-817". The frequency of a measuring field was 30 Hz. The change of the inductance indicates the ferrofluid capillary movement. The magnetic measuring field produced by the solenoid is small enough ($\sim$ 50 A/m), so that its influence on the ferrofluid motion can be neglected. The stationary (maximum reachable) position of wetting front (gas–liquid interface) can be determined by means of sensing inductance coil of small length (5 mm) connected to the measuring bridge (Fig. 1). For this purpose, the sensing coil is moved along the tube with sand to scan the magnetic properties; the abrupt change of the inductance indicates the position of wetting front. The tube used in this experiment is like that used in the experiment on the visual observation of ferrofluid motion in sand.

All the described measurements of the wetting front location have been repeated three times and the mean value was plotted on the graphs. The upper and lower bounds are shown as error bars.

In our experiments we used a kerosene-based ferrofluid with dispersed magnetite nanoparticles of about 10 nm diameter stabilized with oleic acid. The properties of the ferrofluid are: density is $\rho$ = 1000 kg·m$^{-3}$, dynamic viscosity is $\eta$ = 5 mPa·s, initial magnetic permeability is $\mu$ = 1.85, interfacial tension at the ferrofluid–air interface is $\sigma$ = 0.027 N/m. The ferrofluid used was made by OJSC "NIPIgaspererabotka" (Russia). The magnetization curve of ferrofluid is presented in Fig. 4. The magnetization curve has been obtained by the use of a vibrating sample magnetometer (Lake Shore 7410-S). Here the magnetization is plotted as a function of internal magnetic field which has been calculated in a standard way from the applied magnetic field using the demagnetizing factor of the ferrofluid sample shape.

**EXPERIMENTAL RESULTS**

Figure 5 shows the experimental time dependence of the ferrofluid height $h$ in a single cylindrical capillary obtained by the video data processing. The results obtained in zero field and in the presence of non-uniform magnetic field are presented. It can be observed from this figure that the non-uniform external magnetic field with downward directed gradient decelerates the ferrofluid rise and diminishes the maximum height. In contrast, the non-uniform external magnetic field with upward directed gradient accelerates the ferrofluid rise and increases the maximum height. This is due to the action of magnetic net force which is collinear with the magnetic field gradient. In the case of the magnetic field with upward directed gradient the external magnetic field strength at the bottom of the capillary was $H_0$=10 kA/m. In the case of the magnetic field with downward directed gradient the external magnetic field strength at the bottom of the capillary was $H_0$=29 kA/m. In both cases the external magnetic field has changed linearly with height following the law: $H = H_0 + h \cdot \nabla H$, where $\nabla H = 10^6$ A/m$^2$ for the first case, and $\nabla H = -10^6$ A/m$^2$ for the second case.

Magnetic force acting on the ferrofluid depends not only on the magnetic field gradient but also on the magnetic field strength as the magnetization of ferrofluid grows with the magnetic field increasing. Figure 6 shows the experimental dependences of the ferrofluid maximum (static) height $h_{max}$ in a cylindrical capillary on the magnetic field strength at the bottom of capillary $H_0$ obtained by visual observations. As it can be seen, the maximum height decreases with the magnetic field increasing when the magnetic field gradient is down directed and increases when the magnetic field gradient is upward directed. In these experiments the magnetic field gradient has changed only slightly during the magnetic field strength $H_0$ changing as it can be seen from Fig. 3.

Figure 7 shows the experimental time dependence of the ferrofluid height $h$ in a cylindrical tube filled with sand obtained by the video data processing. As is seen from Fig. 7, the character of the non-uniform field influence on the ferrofluid rise in porous medium is analogous to that presented in Fig. 5 and described above. The values of the magnetic field parameters are the same as in the experiments with a single capillary tube without sand.

The visual observations of wetting front motion can be not always effective, especially for the case of materials with visually indistinguishable wet and dry regions. In this case the ferrofluid motion in porous media can be detected by means of magnetic measurements. Figure 8 shows the measured time dependence of the relative inductance $L/L_0$ (here $L_0$ is the initial value of inductance at $t=0$) of solenoid with the sand inside in the absence of external magnetic field. The increase of the solenoid inductance indicates the ferrofluid rise in the sand. The inductance reaches the maximum value $L_{max}$ when the ferrofluid comes up to the top of the solenoid.

The results of scanning the magnetic properties of a tube filled with sand by means of a small sensing inductance coil are presented in Fig. 9. The scanning was performed after the establishment of static position of the wetting front in the absence of external magnetic field. Figure 9 demonstrates the abrupt change of the relative inductance $L/L_d$ at height $h \approx 10$ cm which is the maximum reachable height ($h_{max}$) of ferrofluid in sand in our experiments. Here $L_d$ is coil inductance in the dry region of the sand (at the absence of ferrofluid). Relatively high initial values of the inductance near the tube bottom are associated with the vicinity of the ferrofluid reservoir.

**ANALYSIS OF THE RESULTS**

The momentum balance of a ferrofluid inside a single capillary tube in the presence of a non-uniform magnetic field gives

$$\rho h \ddot{h} = \frac{2\sigma \cos\theta}{R} - \rho g h - \frac{1}{2}\rho \dot{h}^2 - \frac{8\eta}{R^2} h \dot{h} + \mu_0 \int_{H_0}^{H} M(H_i) dH. \qquad (1)$$

Here $R$ is the inner tube radius, $\theta$ is the contact angle formed between solid and liquid, $M(H_i)$ is the magnetization of ferrofluid, $H_i$ is the local magnetic field strength inside the ferrofluid. Since the magnetization of a ferrofluid sample at a given point depends on the internal magnetic field at that point, the demagnetization factor must be used in order to calculate the internal magnetic field from applied (external) magnetic field $H$. In Eq. (1) the individual terms refer to (left to right): the inertia term; the capillary force; the gravity term; the dynamic pressure; the viscous pressure loss; the magnetic force. The following assumptions are made: no friction or inertia effects by displaced air occur; no inertia or entry effects in the liquid reservoir; the flow in the capillary tube is fully developed Poiseuille flow; the contact angle is constant and equal to the static contact angle. By dropping out the last term in Eq. (1) we get the well-known equation for the dynamics of capillary rise in a single capillary [21–23]. Kerosene is possessed of high wetting ability, so we will assume that the contact angle is near zero. This assumption is verified by the comparison of the experimentally obtained static height of ferrofluid in a single capillary in zero field with the corresponding calculated value. It should be noted that the magnetic body force in general has the form: $\mu_0 \int_V M \nabla H dV$, where magnetization $M$ depends on internal magnetic field, and $H$ is the external magnetic field strength [5]. Taking into account that $\nabla H \equiv dH/dh$ in our case, with a little manipulation we get the magnetic force in the form presented in Eq. (1). To perform the calculations the magnetization of ferrofluid can be determined from the experimental magnetization curve (Fig. 4). Magnetic field strength inside the fluid $H_i$ can be computed from the externally applied magnetic field $H$ by the solution of an implicit equation:

$$H_i = H - NM(H_i),\tag{2}$$

where $N$ is the demagnetizing factor, which was set equal to 0.5 for the cylindrical ferrofluid column under study [24]. Here the external magnetic field can be considered to be perpendicular to the capillary. The results of numerical solution of Eqs. (1) and (2) are presented in Fig. 5. As is seen from Fig. 5, the experimental and theoretical results are in a good agreement.

The static balance clearly yields the maximum rise height, i.e. the extended Jurin's law [25]. One can obtain

$$\rho g h_{\max} = \frac{2\sigma \cos\theta}{R} + \mu_0 \int_{H_0}^{H} M(H_i)dH\,.\tag{3}$$

Equation (3) numerical solution is presented in Fig. 6. A qualitative agreement between Eq. (3) and the experimental data can be seen from Fig. 6.

To analyze the ferrofluid rise in a sandy porous medium we will follow the works [21, 23, 26]. The momentum balance can also be given for the capillary rise of ferrofluid in a porous medium; here the viscous term is replaced by the Darcy law

$$\rho h \ddot{h} = \frac{2\sigma \cos\theta}{R} - \rho g h - \frac{1}{2}\rho \dot{h}^2 - \frac{\varphi \eta}{K} h \dot{h} + \mu_0 \int\limits_{H_0}^{H} M\left(H_i\right) dH \,, \tag{4}$$

where $\varphi$ is the porosity of the sand, and $K$ its permeability, $R$ is the pore mean radius. There are different approaches to determine these parameters. Here we will use the experimental results described above to extract the values of the pore structure parameters. The pore radius can be calculated from the maximum obtainable (static) height $h_{max}$. The corresponding calculations show that $R = 55$ μm. The porosity can be calculated from the maximum relative inductance of the solenoid with sand inside using Bruggeman's formula for the magnetic permeability of disperse system [27]:

$$\varphi = \frac{\mu_{ef} - \mu_i}{\mu_e - \mu_i}\left(\frac{\mu_e}{\mu_{ef}}\right)^{1/3} \,, \tag{5}$$

where $\mu_{ef}$ is the effective magnetic permeability of medium, $\mu_i$ is the magnetic permeability of internal phase, $\mu_e$ is the magnetic permeability of external phase. Bruggeman's formula is valid in a wide range of concentrations of disperse system (up to 100 vol. %). In our case the internal phase is presented by sand $\mu_i = 1$, the external phase is presented by ferrofluid $\mu_e \equiv \mu = 1.85$, the effective magnetic permeability is equal to the maximum relative inductance of solenoid $\mu_{ef} = L_{max}/L_0$. Here we assume that the porosity is equal to the saturation, which is only the case if the pore space is completely filled with ferrofluid. The porosity is calculated to be $\varphi = 0.38$ in our case. The permeability $K$ can be obtained by fitting the calculated according to Eq. (4) ferrofluid rise curve to experimental values (Fig. 7). Numerical solution of Eq. (4) at the absence of external magnetic field and comparison with the corresponding experimental values give $K = 1.5 \cdot 10^{-11}$ m$^2$. Using the obtained values of the pore structure parameters we can also calculate the ferrofluid rise curve in the presence of external non-uniform magnetic field (Fig. 7). As is seen from Fig. 7, in the case of downward directed magnetic field gradient the difference between the experimental and theoretical results is notable, on the other hand in the case of upward directed magnetic field gradient the experimental and theoretical results are in a good agreement (the difference is no more than 10%). When the ferrofluid penetrates into the porous medium the strong magnetic field can influence the medium microstructure. It is known that nonmagnetic particles immersed in the ferrofluid (sand grains in our case) can be arranged along the magnetic field lines. The force of interaction between non-magnetic particles in ferrofluid leading to the structure formation is proportional to the magnetic field strength, not to the magnetic field gradient. When the magnetic field gradient is downward directed, the region of penetration of ferrofluid into the porous medium is located in a strong magnetic field. So the structure formation will be more intensive in this case than in the case of upward directed gradient, although the absolute values of field gradient are the same in the both cases. This can lead to the resulting difference of experimental and theoretical results when the magnetic field gradient is down-directed.

In the above calculations it has been assumed that the magnetic field does not change the shape of a ferrofluid meniscus. Let us estimate the validity of this assumption. The magnetic pressure acting at some point on the ferrofluid interface can be calculated as $P_m = \frac{1}{2}\mu_0 (\mu - 1) H_\tau^2$, where $H_\tau$ is the tangential component of the magnetic field strength. Supposing that the meniscus shape is spherical, the capillary pressure can be calculated as $P_c = 2\sigma/R$. Let $H_\tau = 10$ kA/m. The corresponding calculation shows that the magnetic pressure makes less than 0.1 of the capillary pressure for the porous medium of sand, and less than 0.2 for the single capillary tube studied. It can be concluded that the change of the meniscus shape caused by the magnetic field should take place but it will be not so notable. The presented estimation demonstrates the possible value of error contributed by the neglect of the magnetic field effect on the ferrofluid meniscus shape.

Consider the change of the solenoid inductance during the ferrofluid capillary rise. The inductance can be presented as a sum of inductances of wetted and dry regions: $L = L_1 + L_2$. We have: $L_1 = \mu_0\mu_{ef}n^2hS$, $L_2 = \mu_0 n^2 lS$, where $\mu_0$ is the magnetic constant $\mu_0 = 4\pi \cdot 10^{-7}$ H/m, $n$ is the number of turns per unit length, $S$ is the solenoid cross-section area, $l$ is the length of dry region. Taking into account that: $l_0 = h + l$, $L_0 = \mu_0 n^2 l_0 S$, $L_{max} = \mu_0\mu_{ef}n^2 l_0 S$, it is easy to obtain

$$h(t) = l_0 \frac{L(t) - L_0}{L_{max} - L_0}, \tag{6}$$

where $l_0$ is the solenoid total length. Figure 10 shows the ferrofluid rise curve calculated according to Eq. (6) from the data presented in Fig. 8 in comparison with the corresponding results obtained by visual observations in zero field. As is seen, the results obtained by means of magnetic measurements are in a good agreement with the results of visual observations (the difference is no more than 10%).

## DISCUSSION AND CONCLUSIONS

The study of liquid capillary rise in a cylindrical column of porous material is used to extract the pore structure parameters [20, 21, 23, 26]. The results described above allow us to propose a modified method of evaluation of porous material parameters based on the use of magnetic measurements. Utilizing the ferrofluid as a liquid which capillary ascension is analyzed, we can extract all porous material parameters from the results of measurements of the inductances of the solenoid and the small censing coil. This method can be especially useful for the case of materials with visually indistinguishable wet and dry regions. For validation of the obtained results we have applied different experimental methods to determine the pore structure parameters of the sand. The porosity has been determined from the weight of a completely wetted specimen. The permeability has been determined by applying a defined pressure difference to a porous structure and measuring the resulting flow rate. For this purpose we used common procedure and equipment described, e.g. in [28]. A comparison of the results showed that the differences are no more than 10%. It should be noted that the proposed method is limited to the cases when the discrete nature of ferrofluid can be neglected, i.e. when the pore sizes are large compared to the ferrofluid nanoparticle sizes. Another limitation is the validity of Eq. (6), which is only the case if the rise height can be comparatively large. The ability of the use of ferrofluids based on the various types of liquids makes some advance of facilities of the proposed method.

Thus, in the present study we have investigated the linear capillary transport of ferrofluid in a single cylindrical tube and in a porous medium. Our experimental investigation revealed new aspects of the external magnetic field influence on the ferrofluid capillary flow. The obtained experimental results have been theoretically analyzed. A comparison of experimental and theoretical results shows that they are in a qualitative agreement. It should be noted that several aspects of capillary flow have not been considered in the performed analyses. These include the local acceleration of the liquid below the tube, viscous losses below the tube, convective flow losses at the entrance of the tube, evaporation effects, ferrofluid meniscus shape change, pore shape and size distribution, etc.

**ACKNOWLEDGMENT**

This work was supported by the grant of the President of the Russian Federation No. MK-5801.2015.1 and also by the Ministry of Education and Science of the Russian Federation in the framework of the base part of the governmental ordering for scientific research works (order No 2014/216, project 2479).

**NOMENCLATURE**

| | |
|---|---|
| $H$ | external (applied) magnetic field strength, A/m |
| $H_0$ | external magnetic field strength at the bottom of capillary, A/m |
| $H_i$ | internal magnetic field strength inside the ferrofluid, A/m |
| $H_\tau$ | tangential component of the external magnetic field strength, A/m |
| $h$ | ferrofluid height, m |
| $h_{max}$ | ferrofluid maximum (static) height, m |
| $K$ | sand permeability, $m^2$ |
| $L$ | solenoid (or sensing coil) inductance, H |
| $L_0$ | initial value of solenoid inductance at $t$=0, H |
| $L_1$ | inductance of wetted region of solenoid, H |
| $L_2$ | inductance of dry region of solenoid, H |
| $L_d$ | sensing coil inductance in the dry region of sand, H |
| $L_{max}$ | maximum value of solenoid inductance, H |
| $l$ | length of dry region of sand column inside solenoid, m |
| $l_0$ | solenoid total length, m |
| $M(H_i)$ | magnetization of ferrofluid, A/m |
| $N$ | demagnetizing factor |
| $n$ | number of turns per unit length of solenoid, $m^{-1}$ |
| $P_c$ | capillary pressure, Pa |
| $P_m$ | magnetic pressure, Pa |
| $R$ | inner capillary tube radius or the pore mean radius, m |
| $S$ | solenoid cross-section area, $m^2$ |
| $t$ | time, s |
| $V$ | volume of ferrofluid, $m^3$ |
| $\eta$ | ferrofluid dynamic viscosity, Pa·s |
| $\theta$ | contact angle formed between solid and liquid, rad |
| $\mu$ | ferrofluid initial magnetic permeability |
| $\mu_0$ | magnetic constant, $4\pi \cdot 10^{-7}$ H/m |
| $\mu_{ef}$ | effective magnetic permeability of medium |
| $\mu_e$ | magnetic permeability of external phase |
| $\mu_i$ | magnetic permeability of internal phase |
| $\rho$ | ferrofluid density, $kg/m^3$ |
| $\sigma$ | interfacial tension at the ferrofluid–air interface, N/m |
| $\varphi$ | sand porosity |

**REFERENCES**

[1] Van Brakel, J., and Heertjes, P.M., 1975, "Capillary rise in porous media," Nature, **254**, pp. 585–586.

[2] Alava, M., Dubé, M., and Rost, M., 2004, "Imbibition in disordered media," Adv. Phys., **53**(2), pp. 83–175.

[3] Nia, S.F., and Jessen, K., 2015, "Theoretical analysis of capillary rise in porous media," Transport Porous Med., **110**(1), pp. 141–155.

[4] Vafai, K., (Ed.), 2015, *Handbook of Porous Media*, CRC Press, Boca Raton.

[5] Rosensweig, R.E., 1985, *Ferrohydrodynamics*, Cambridge University Press, Cambridge.

[6] Sadrhosseini, H., Sehat, A., and Shafii, M.B., 2016, "Effect of magnetic field on internal forced convection of ferrofluid flow in porous media," Exp. Heat Transfer, **29**(1), pp. 1–16.

[7] Bashtovoi, V., Kuzhir, P., and Reks, A., 2002, "Capillary ascension of magnetic fluids," J. Magn. Magn. Mater., **252**(1), pp. 265–267.

[8] Lee, C.-P., Chang, H.-C., and Lai, M.-F., 2011, "Magnetocapillary phenomenon affected by magnetic films," J. Appl. Phys., **109**, 07E310.

[9] Bashtovoi, V., Bossis, G., Kuzhir, P., and Reks, A., 2005, "Magnetic field effect on capillary rise of magnetic fluids," J. Magn. Magn. Mater., **289**, pp. 376–378.

[10] Borglin, S.E., Moridis, G.J., and Oldenburg, C.M., 2000, "Experimental studies of the flow of ferrofluid in porous media," Transport Porous Med., **41**(1), pp. 61–80.

[11] Oldenburg, C.M., Borglin, S.E., and Moridis, G.J., 2000, "Numerical simulation of ferrofluid flow for subsurface environmental engineering applications," Transport Porous Med., **38**(3), pp. 319–344.

[12] Ivanov, A.B., and Taktarov, N.G., 1990, "A study into the filtration of magnetic fluids," Magnitnaya Gidrodinamika, **26**(3), pp. 390–392.

[13] Qin, Y., and Chadam, J., 1995, "A nonlinear stability problem of ferromagnetic fluids saturating a porous medium with inertial effect," Appl. Math. Lett., **8**(2), pp. 25–29.

[14] Vaidyanathan, G., and Sekar, R., 1991, "Ferroconvective instability of fluids saturating a porous medium," Int. J. Eng. Sci., **29**(10), pp. 1259–1267.

[15] Verma, A.P., and Rajput, A.K., 1987, "Instabilities in displacement processes through porous media with magnetic fluid," J. Magn. Magn. Mater., **65**(2–3), pp. 330–334.

[16] Taktarov, N.G., 1981, "Convection of magnetizable fluids in porous media," Magnitnaya Gidrodinamika, **17**(4), pp. 333–335.

[17] Larachi, F., and Desvigne, D., 2006, "Magnetoviscous control of wall channeling in packed beds using magnetic nanoparticles – Volume average ferrohydrodynamic model and numerical simulations," Chem. Eng. Sci., **61**(5), pp. 1627–1657.

[18] Larachi, F., and Desvigne, D., 2007, "Ferrofluid induced-field effects in inhomogeneous porous media under linear-gradient d.c. magnetic fields," Chem. Eng. Process., **46**(8), pp. 729–735.

[19] Larachi, F., and Desvigne, D., 2007, "Ferrofluid magnetoviscous control of wall channelling in porous media," China Particuology, **5**(1–2), pp. 50–60.

[20] Lago, M., and Araujo, M., 2001, "Capillary rise in porous media," J. Colloid Interface Sci., **234**(1), pp. 35–43.

[21] Fries, N., and Dreyer, M., 2008, "An analytic solution of capillary rise restrained by gravity," J. Colloid Interface Sci., **320**(1), pp. 259–263.

[22] Fries, N., and Dreyer, M., 2008, "The transition from inertial to viscous flow in capillary rise," J. Colloid Interface Sci., **327**(1), pp. 125–128.

[23] Fries, N., and Dreyer, M., 2009, "Dimensionless scaling methods for capillary rise," J. Colloid Interface Sci., **338**(2), pp. 514–518.

[24] Osborn, J.A., 1945, "Demagnetizing factors of the general ellipsoid," Phys. Rev., **67**(11, 12), pp. 351–357.

[25] de Gennes, P.-G., Brochard-Wyart, F., and Quere, D., 2004, *Capillarity and Wetting Phenomena: Drops, Bubbles, Pearls, Waves*, Springer-Verlag, New York.

[26] Fries, N., Odic, K., Conrath, M., and Dreyer, M., 2008, "The effect of evaporation on the wicking of liquids into a metallic weave," J. Colloid Interface Sci., **321**(1), pp. 118–129.

[27] Bruggeman, D.A.G., 1935, "Berechnung verschiedener physikalischer konstanten von heterogenen substanzen. I. Dielektrizitätskonstanten und leitfähigkeiten der mischkörper aus isotropen substanzen," Ann. Phys. **416**(7), pp. 636–664.

[28] Tiab, D., Donaldson, E.C., 2012, *Petrophysics*, Elsevier, Oxford.

## Figure Captions List

Fig. 1     Experimental setup

Fig. 2     Microscopic image of the grains of the sand sample used in the experiments

Fig. 3     Example of the dependences of the magnetic field strength on the distance from the electromagnet poles at different values of the field strengths in the immediate vicinity of the poles $H_0$.

Fig. 4     Magnetization curve of the ferrofluid used in experiments

Fig. 5     Time dependence of the ferrofluid height in a single cylindrical capillary for different orientations of the magnetic field gradient. Dots represent experimental data. Corresponding calculations are shown by solid lines

Fig. 6     Maximum reachable height of ferrofluid in a single cylindrical capillary versus the magnetic field strength at the capillary bottom. Dots are experiments; lines are calculations

Fig. 7     Time dependence of the ferrofluid height in a sandy porous medium for different orientations of the magnetic field gradient. Dots are experiments; lines are calculations

Fig. 8     Time dependence of the relative inductance of the solenoid filled with sand during the ferrofluid capillary rise

Fig. 9     The relative inductance of the sensing coil as a function of its location (height) on a tube filled with sand after the establishment of static position

of wetting front

Fig. 10    Time dependence of the ferrofluid height in a sand column obtained by means of magnetic measurements (calculated according to Eq. (6) from the data presented in Fig. 8) in comparison with the results of visual observations presented in Fig. 7

**Fig. 1**

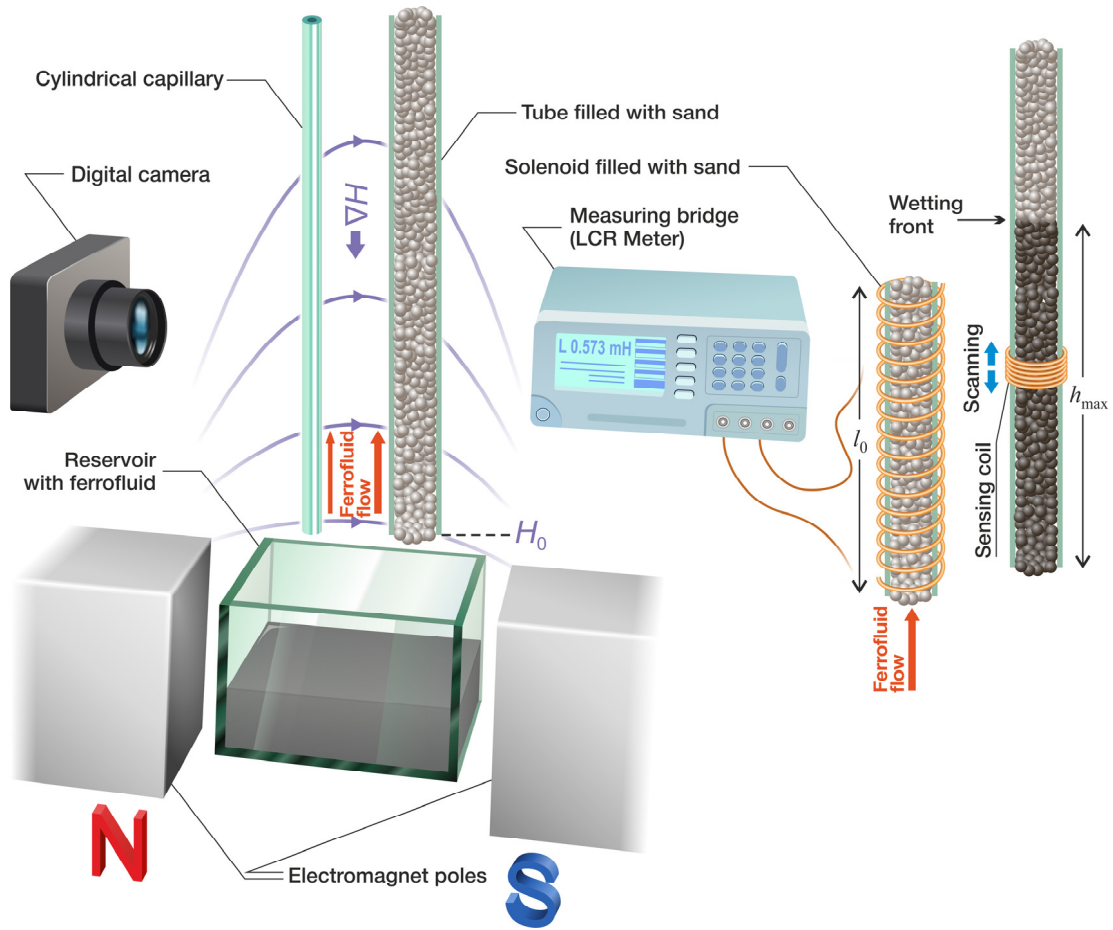

**Fig. 2**

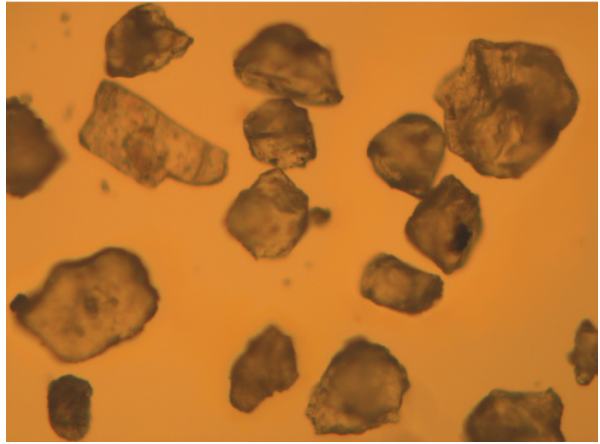

**Fig. 3**

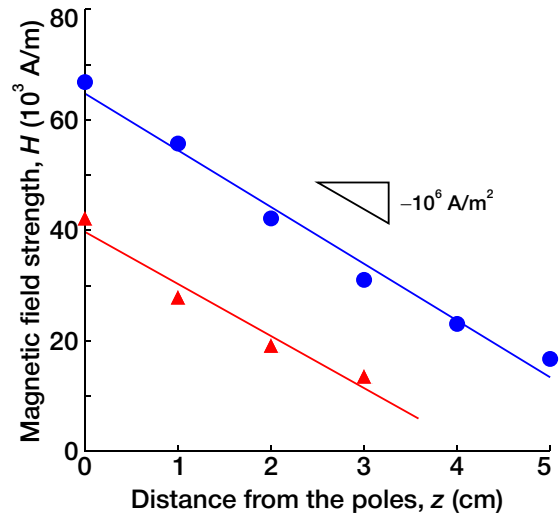

**Fig. 4**

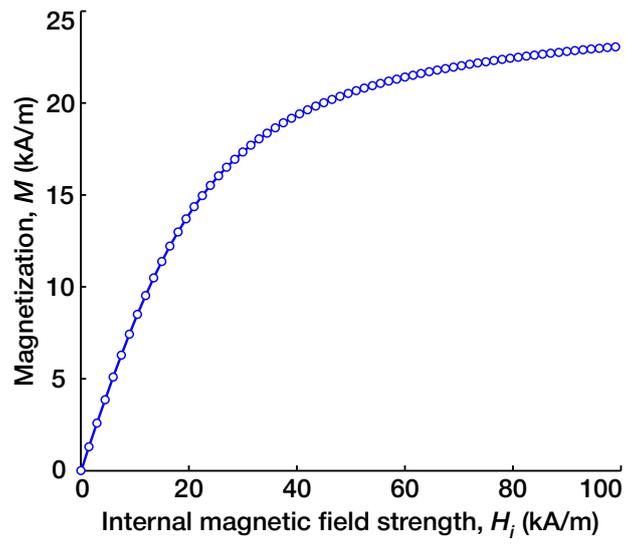

**Fig. 5**

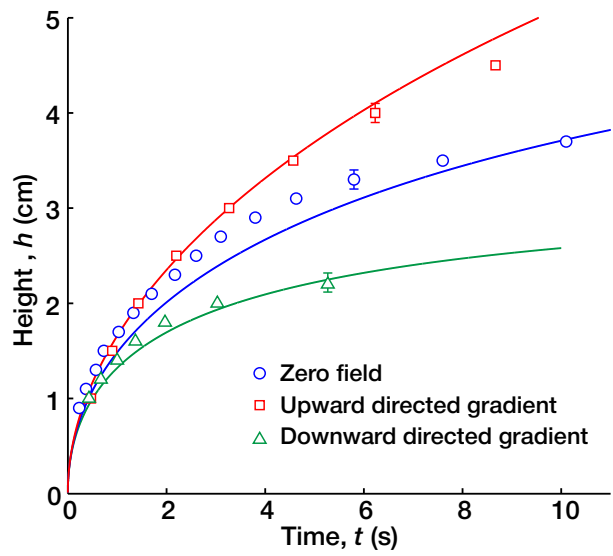

**Fig. 6**

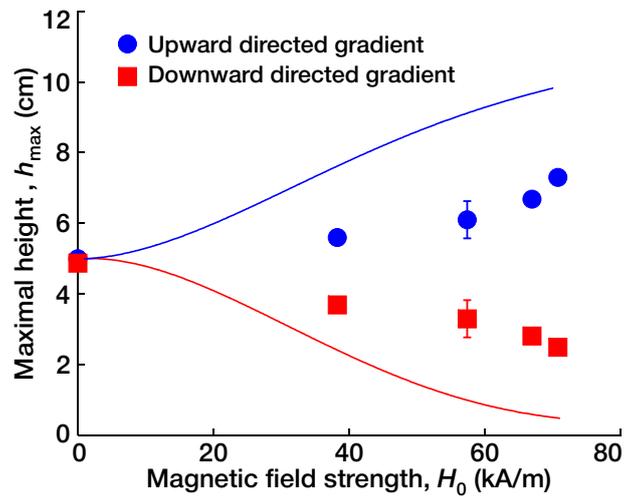

**Fig. 7**

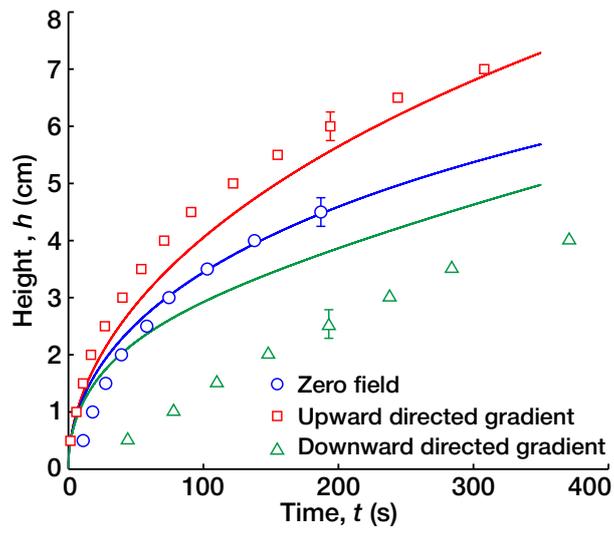

**Fig. 8**

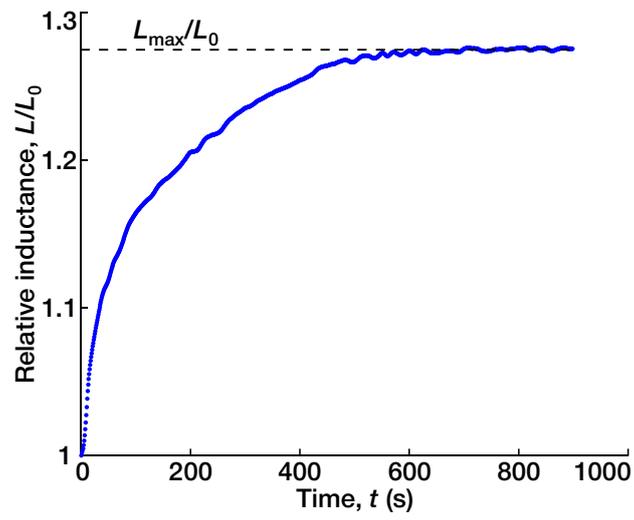

**Fig. 9**

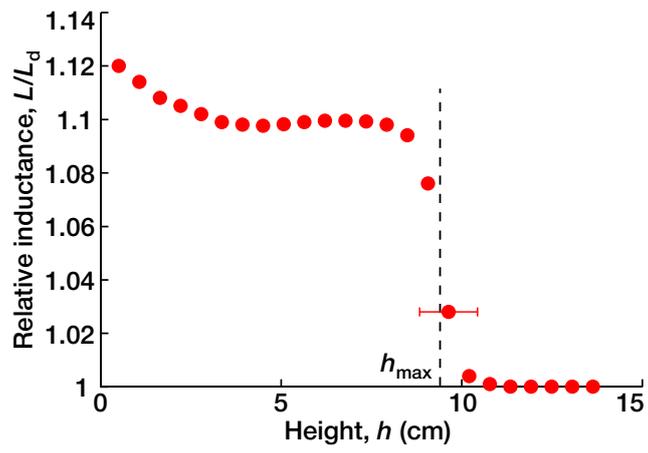

**Fig. 10**

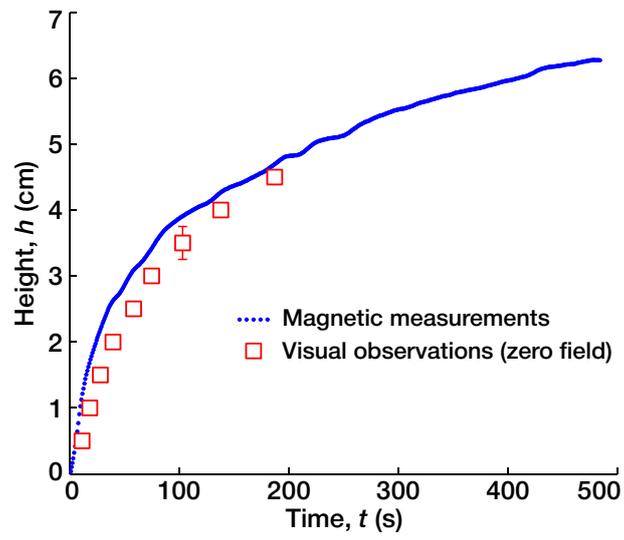